**Title: Ultrahigh Resolution Spectroscopy Across the Visible to Infrared Spectrum Using Multi-Mode Interference in a Compact Tapered Fiber**

**Authors:**

| | |
|---|---|
| Noel H. Wan[1,2]† | nhw2109@columbia.edu |
| Fan Meng[1,3]† | fm2450@mit.edu |
| Ren-Jye Shiue[1] | tedshiue@mit.edu |
| Edward H. Chen[1] | ehchen@mit.edu |
| Tim Schröder[1] | schroder@mit.edu |
| Dirk Englund[1] | englund@mit.edu |

†These authors contributed equally to this work

**Correspondence**:
Dirk Englund
Email: englund@mit.edu;
Address: Room 36-591, 77 Massachusetts Avenue, Cambridge MA, 02139-4307, USA
Telephone: +16173247014

**Affiliations:**

1. Department of Electrical Engineering and Computer Science, Massachusetts Institute of Technology, Cambridge, Massachusetts 02139, USA.

2. Department of Physics, Columbia University, New York, New York 10027, USA

3. State Key Laboratory of Information Photonics and Optical Communications, Beijing University of Posts and Telecommunications, Beijing 100876, China

**Abstract**:

Optical spectroscopy is a fundamental tool in numerous areas of science and technology. Much effort has focused on miniaturizing spectrometers, but thus far at the cost of high spectral resolution and broad operating range. Here, we describe a compact spectrometer without this trade-off. The device relies on imaging multi-mode interference from leaky modes along a highly multimode tapered optical fiber, resulting in spectrally distinguishable images that form a basis for reconstructing an incident light spectrum. This tapered fiber multimode interference spectrometer enables the acquisition of broadband spectra in a single camera exposure with a measured resolution of 40 pm in the visible spectrum and 10 pm in the infrared spectrum (corresponding to quality factors of $10^4 - 10^5$), which are comparable to the performance of grating spectrometers. Spectroscopy from 500 nm to 1600 nm is demonstrated, though operation across the entire transparency window of silica fibers is possible. Multimode interference spectroscopy of leaky modes is suitable in a variety of device geometries, including planar waveguides in a broad range of transparent materials. We anticipate that this technique will greatly expand the availability of high-performance spectroscopy in a wide range of applications.

**INTRODUCTION**

Since its 19th century origins with Walleston and Frauenhofer[1], modern spectroscopy has become an essential tool for scientific research, including analytical chemistry, biochemical sensing[2,3], material analysis[4], optical communication[5,6], and medical applications[7]. Modern grating spectrometers project the spectral components of light onto a photodetector array, and some commercially available systems[8,9] are able to achieve very high resolving powers exceeding $Q>10^7$ as quantified by the quality factor $Q = \frac{\omega_0}{\delta\omega}$, where $\delta\omega$ is the spectral resolution at frequency $\omega_0$. However, high-resolution grating spectrometers are inevitably bulky because their spectral resolution scales inversely with optical path length. Fourier Transform spectrometers can be more compact in size, but their scanning interferometric configuration makes them slow for large bandwidth applications.

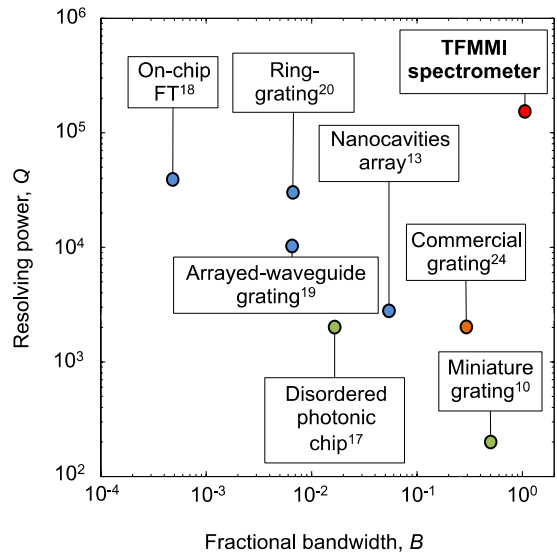

**Figure 1** Figures of merit of several compact spectrometers employing detector arrays, including the tapered fiber multimode interference (TFMMI) spectrometer that is described in this work. The TFMMI spectrometer achieves a resolving power of $Q > 10^4$ ($Q > 10^5$) in the visible (near-infrared) spectrum, and bandwidth $B \sim 1$. For the purpose of comparison, we include a commercial single-grating spectrometer[24]. Table 1 in the Supplementary Information provides more details about these spectrometers.

In recent years, there has been a strong effort to develop novel spectrometers that are compact and have high throughput and resolving powers[5,10-23]. In addition to miniature grating spectrometers[10], resonant structures such as nanocavities[11-13] have been employed to separate spectral components into unique detectors. This research has produced millimeter-scale spectrometers with high resolving powers of $Q>10^4$, but with limited fractional bandwidth $B = \frac{\omega_{max} - \omega_{min}}{(\omega_{max} + \omega_{min})/2}$. Another class of compact spectrometers relies on imaging the speckle patterns of a photonic bandgap fiber bundle[14] or a multimode optical fiber[15,16]. However, these spectrometers are centimeters to meters long, which can result in instability of the interference pattern due to environmental fluctuations.

Recently, a silicon-based spectrometer[17] that measures the multimode transmission profile from a disordered photonic crystal structure achieved a $Q$ comparable to typical grating spectrometers, but it requires very precise coupling to a sub-wavelength input waveguide and has a small bandwidth from 1500 nm to 1525 nm. A high-resolution Fourier Transform spectrometer[18] has also been demonstrated on-chip, but its narrow free spectral range limits its operation bandwidth to 0.75 nm at 1550 nm. Figure 1 lists various compact spectrometers in terms of their resolving power, $Q$, and fractional bandwidth, $B$. As seen from Fig. 1, although some of these recently developed compact spectrometers can already match the spectral resolution of grating spectrometers, it has so far been elusive to achieve high resolving power *and* broad operating bandwidth in one compact device.

Here, we introduce a monolithic compact spectrometer based on multimode interference (MMI) that simultaneously achieves high resolving power and bandwidth in a simple-to-couple and compact device. The device consists of a tapered multi-mode fiber, as illustrated in Fig. 2a. A non-adiabatic taper angle causes light at frequency $\omega$ to couple to a large number of spatial modes with different propagation constants, resulting in a unique interference profile that can be imaged as the 'fingerprint', or basis state, of a given frequency $\omega$. Specifically, in the tapered optical fiber, the electric field varies in the propagation direction (*z*) as

$$E(z,t) = \exp(i\omega t) \sum_{l,m} E_{0,l,m} \exp(i\beta_{l,m}(z)z)$$

where $\beta_{l,m}(z)$ are the phase propagation constants of the fiber Bessel modes, plotted in Fig. 2b. The tapering of the fiber additionally modifies this interference along the taper length, as reflected in the *z*-dependent propagation constant $\beta_{l,m}(z)$, while also introducing a controllable mode leakage. Imaging the interference pattern $I = |E(x,y,z)|^2$ with a camera provides a system with exceptional spectroscopic attributes that enable order-of-magnitude improvements over state of the art spectrometers (Fig. 1).

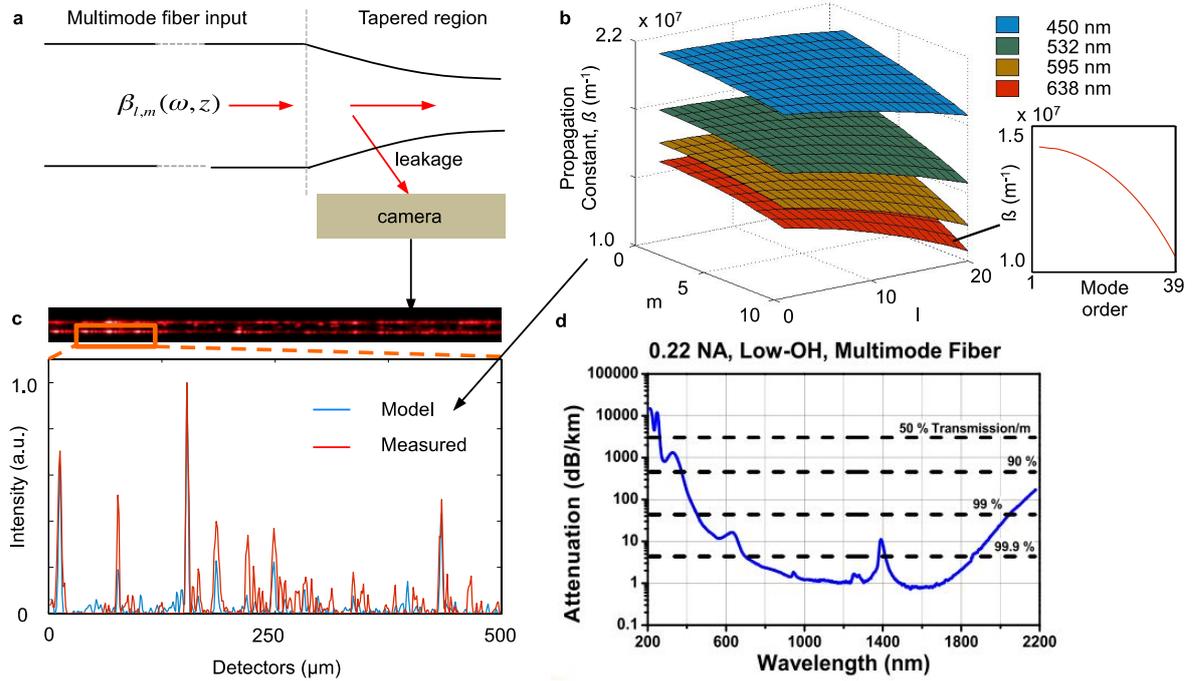

**Figure 2** (**a**) Schematic of the TFMMI spectrometer, consisting of a tapered fiber imaged on a commercial camera. (**b**) Isoplanes of propagation constants at four frequencies corresponding to $\lambda$ = 450 nm, 532 nm, 595 nm and 638 nm. *Inset*: propagation constant as a function of mode order. (**c**) Optical image of the device when a 638 nm laser is sent through the tapered fiber. The intensity distribution is extracted from one MMI section for the purpose of illustration, showing qualitative agreement between experiment and model. (**d**) Transmission spectrum of the fiber (Thorlabs AFS105/125y)[25]. We anticipate an operating range equivalent to the supported wavelengths from 400 nm to 2400 nm.

**MATERIALS AND METHODS**

The fabrication of the devices relies on tapering a multimode optical fiber (Thorlabs AFS105/125y) using the flame-brush technique[26] with taper lengths ~5 mm. The MMI patterns were imaged in a central ~1.5 mm stretch of the fiber, using a microscope as illustrated in Fig. 2a. A fiber polarization controller before the TFMMI spectrometer maximizes the coupling efficiency. Figure 3a and 3b show the interference patterns obtained for different input wavelengths of 450 nm, 532 nm, 595 nm and 638 nm in the visible spectrum and 1560 nm, 1561 nm and 1562 nm in near-infrared, respectively. Different laser frequencies correspond to clearly resolvable MMI images.

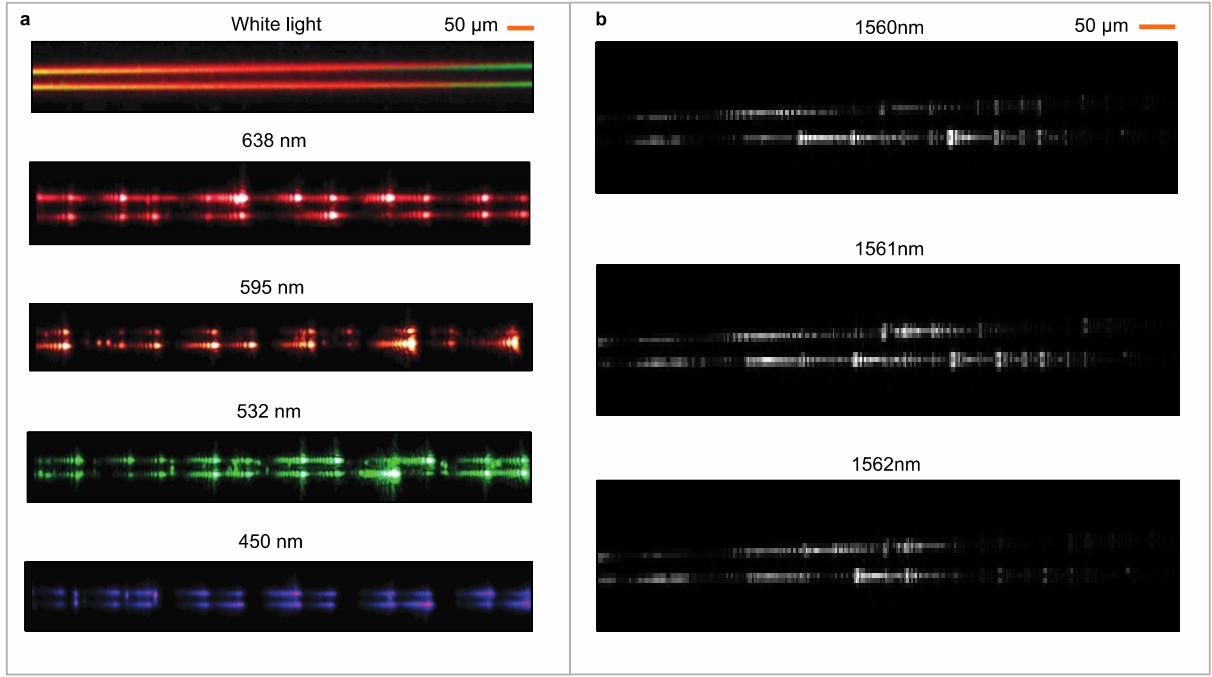

**Figure 3 (a)** Optical images of white light, 638 nm, 595 nm, 532 nm and 450 nm monochromatic fields through the tapered fiber. **(b)** Monochrome CCD images showing the spatial intensity distribution for wavelengths of 1560 nm, 1561 nm and 1562 nm. Scale bar: 50 µm.

We first characterized the TFMMI spectrometer using narrow laser fields (linewidths < 10 MHz) tuned from 634.800 nm to 639.400 nm in steps of 2 pm, and from 1500.000 nm to 1580.000 nm in steps of 1 pm. The resulting images with pixel dimensions $X \times Y$ are concatenated to produce vectors $\vec{u}$ of length $L = X \times Y$. The vectors $\vec{u}$ resulting from the intensity distributions are normalized and stored in an $L \times N$ 'calibration matrix' $\Lambda$, where $N$ is the number of scanned wavelengths. Because a single-mode fiber was used to direct light into the device, the wavelength-associated fingerprints that constitute this calibration matrix are independent of the actual input light profile. For arbitrary incident light from a single mode fiber, the insertion loss from single mode to multimode fibers is only 0.1 dB via a standard fiber-optic connector. In practice, a single-mode fiber could also be spliced to the tapered multimode fiber to provide a fixed connection.

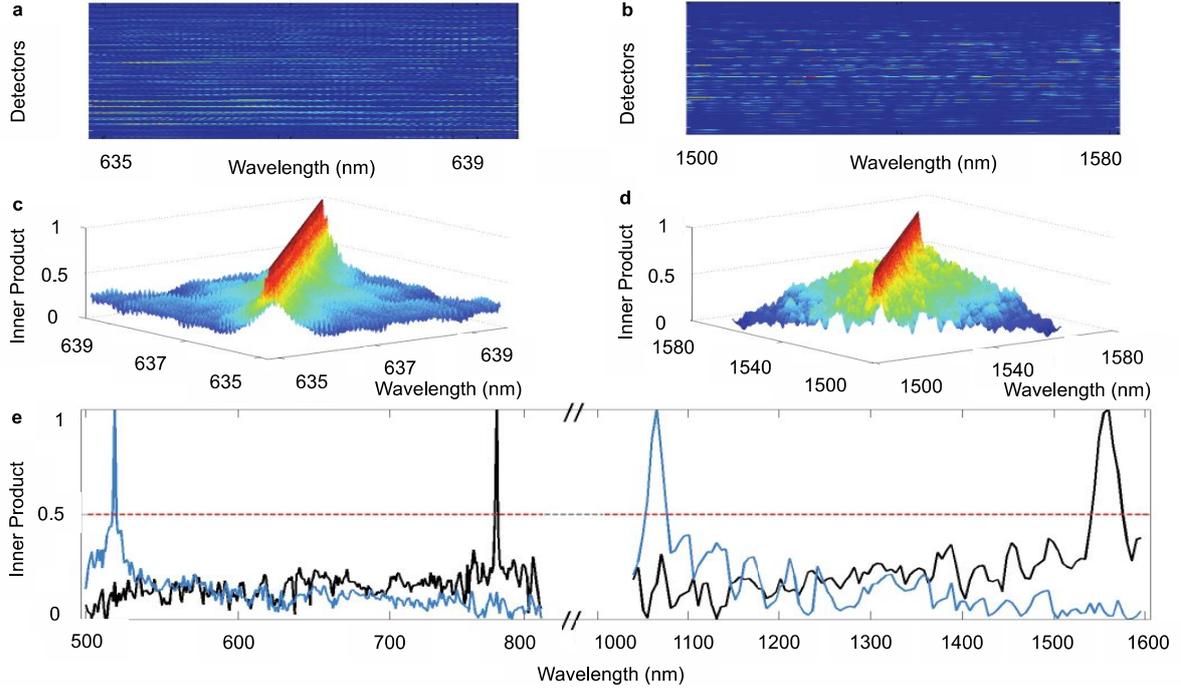

**Figure 4** Frequency states corresponding to intensity fingerprints describe a spectral-to-spatial map. **(a, b)** Response matrix showing the intensity distributions of different wavelengths through the fiber from 634.800 nm to 639.400 nm and from 1500.000 nm to 1580.000 nm, respectively. **(c, d)** Monotonically decreasing inner products $\vec{u}_i \cdot \vec{u}_j$ showing decreasing overlap of spatial distribution of intensity in the maps in **(a)** and **(b)**, respectively. **(e)** The inner products between wavelengths from 500 nm to 800 nm and from 1040 nm to 1595 nm decay to ~$10^{-1}$ levels, showing that no two distinguishable wavelengths have the same associated fingerprint.

Figure 4a and 4b show the normalized intensity distribution along the fiber for visible and infrared wavelengths, respectively. Using the maps in Fig. 4a and 4b as our calibration matrices, we computed every pairwise permutation of inner products. For an ideal TFMMI of long length, after normalization, $\vec{u}_i \cdot \vec{u}_i = 1$ and $\vec{u}_i \cdot \vec{u}_j = 0$ for all $i \neq j$. In practice, however, there are non-trivial overlapping spatial channels between these states due to the finite length of the fiber, but as seen from Fig. 4c and 4d, the inner products $\vec{u}_i \cdot \vec{u}_j$ monotonically decrease to isolation levels of ~$10^{-1}$ as the wavelength separation $|\vec{u}_j - \vec{u}_i|$ increases. This well-behaved complete basis set is desirable for high-resolution and broadband spectroscopy.

Due to the broad transparency window of the silica fiber, the TFMMI spectrometer supports a broad wavelength range from 400 nm up to 2400 nm in principle (Fig. 1d). We investigated this spectroscopy range by extending our previous analysis to 500 nm< $\lambda$ < 800 nm and 1100 nm< $\lambda$ <1600 nm. Figure 4e again indicates a nearly orthonormal basis in these wavelength ranges.

Broad-range spectroscopy can be performed for any input spectrum after calibrating with a full-rank matrix $\Lambda$: Given any arbitrary measured response signal $\vec{\Psi}$, one recovers the amplitude vector constituting the signal's spectrum, $\vec{s} = (s_1, s_2, ..., s_N)$ by left-inversion using the Moore-Penrose pseudo-inverse matrix, $\Lambda^+$ to obtain the least-squares solution $\vec{s} = \Lambda^+ \vec{\Psi}$. This least-square solution is desirable because it applies in situations when $\vec{\Psi} \notin span\{\vec{u}_i\}_{i=1}^N$, which is relevant in spectroscopy due to the continuity of spectra and the presence of experimental noise. Thus, with the least-squares solution, we can impose the condition that $s_j > 0$ for all $j$ and solve the more general nonlinear minimization problem $\|\Lambda \vec{s} - \vec{\Psi}\|_2$ [15-17, 27] to extract the spectral component $\vec{s}$ more accurately.

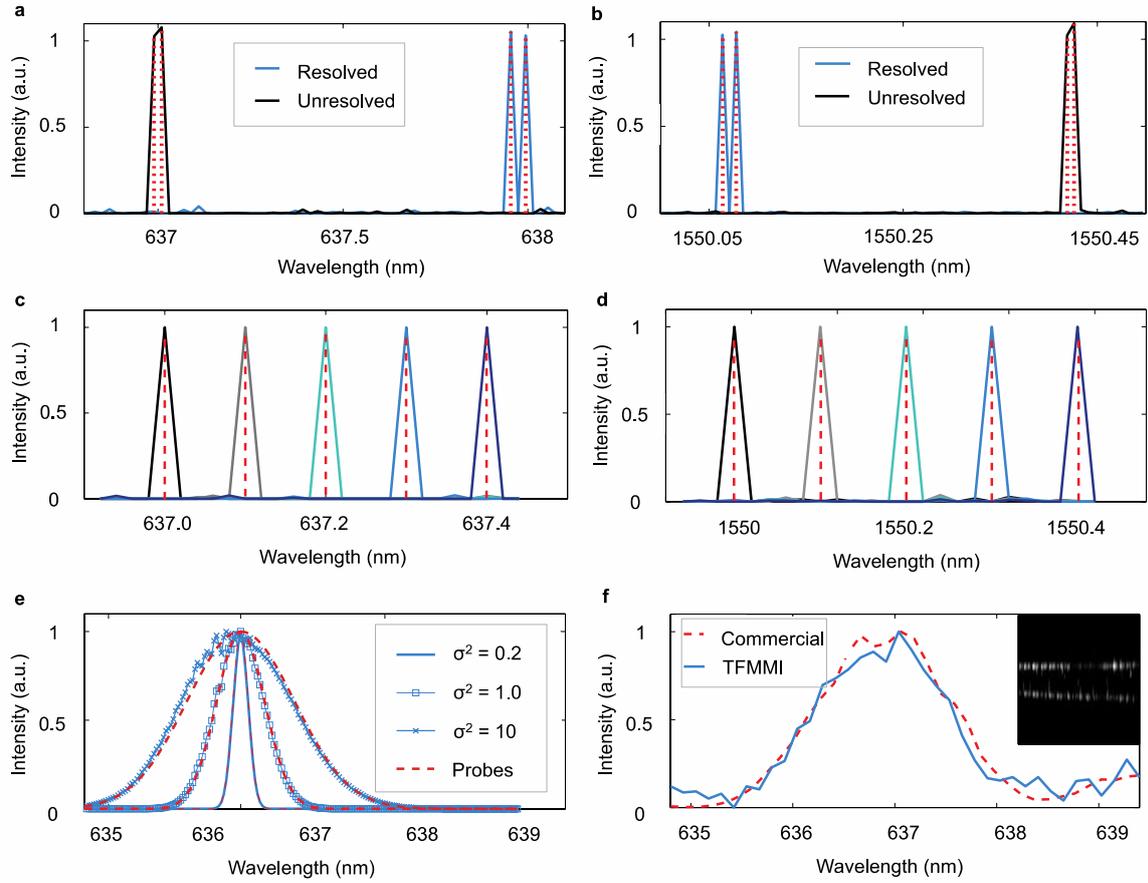

**Figure 5** Performance of TFMMI spectrometer. **(a)** Sharply peaked input wavelengths of 637.96 nm and 638.00 nm are clearly resolved while input wavelengths of 637.00 nm and 637.02 nm are not fully resolved, indicating a resolution of the TFMMI of 40 pm at 638 nm. **(b)** Sharply peaked input wavelengths of 1550.070 nm and 1550.080 nm are clearly resolved while input wavelengths of 1550.410 nm and 1550.415 nm are not fully resolved; thus the resolution of the device is 10 pm at 1550 nm. **(c, d)** Reconstruction of multiple narrow spectra in the visible and in the infrared, respectively. **(e)** Reconstruction of Gaussian-enveloped spectra with different variances. **(f)** Reconstructed spectrum of a filtered broadband signal using a device resolution equivalent to that of the commercial spectrometer used, showing good agreement with the spectrum obtained from this spectrometer.

**RESULTS AND DISCUSSION**

To determine the spectral resolution of the TFMMI spectrometer, we measured the intensity profiles of two nearby input wavelengths at various power levels. Images were acquired sequentially and summed to reflect superposition. Over tens-of-milliseconds integration times of these experiments, there is no interference between laser fields for laser separations of more than 1 pm (0.8 GHz). Therefore, adding sequential measurements at different frequencies is justified (though additional background noise is introduced compared to a single exposure). The separation between the pair of wavelengths was gradually increased to find the wavelength separation at which the reconstruction algorithm could resolve them. This is the case at $\Delta\lambda = $ 40 pm in the visible regime (Fig. 5a) and $\Delta\lambda = $ 10 pm in the IR regime (Fig. 5b), providing the upper bounds on the TFMMI resolution in these spectral ranges.

In Fig. 5c and 5d, the TFMMI was used to reconstruct multiple narrow spectra in the visible and in the infrared, respectively. We also tested the TFMMI spectrometer with three simulated Gaussian spectrum input probe fields of increasing widths. As shown in Fig. 5e, the root-mean-square-deviation of the reconstructed amplitudes from the Gaussian probe with a 10 nm variance is 0.0257.

As an application of this spectrometer, we introduced a broadband signal that was within the calibrated range of our device. We generated this signal by filtering a supercontinuum light source (SuperK, NKT Photonics), which resulted in a probe light of ~1.5 nm bandwidth around 637 nm. The probe light was then directed to our TFMMI spectrometer and to a commercial 0.75 m grating spectrometer[22]. As shown in Fig. 5f, the reconstructed spectrum of the TFMMI spectrometer matches the result from the commercial spectrometer.

The performance of such devices can be improved by optimizing the waveguide or fiber width. To address this question, we analyzed two regions along the TFMMI spectrometer with diameters of 15 µm and 30 µm using two spectral ranges: 636.8-639.4 nm and 768-778 nm. Figure 6a shows that the inner products from these two regions fall to 0.5 after and ±0.8 nm and ±1.0 nm from the central wavelength, respectively. This is because the greater number of modes in the thicker section causes an MMI pattern that is more sensitive to wavelength changes. In addition, because the taper width falls off approximately exponentially[28] with $z$, the steeper profile of the thicker region also means the MMI leakage is more spatially dependent along the fiber and leads to more non-uniformity in intensity distribution for a particular wavelength. Because the device performance scales with length, the taper profile can be suitably engineered to optimize both taper angle and length to improve the performance of the device. In addition, the refractive indices and materials can also be further optimized to increase the MMI contrast, the operational range of the TFMMI spectrometer.

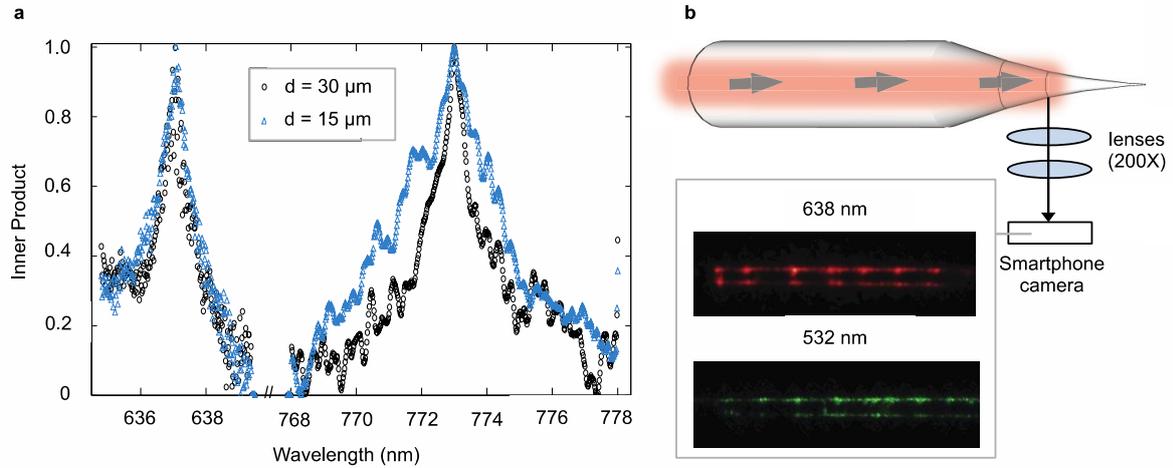

**Figure 6 (a)** Two regions along the TFMMI spectrometer with diameters of 30 µm and 15 µm. The inner products of visible wavelengths from these two regions fall to 0.5 after and ±0.8 nm and ±1.0 nm from the central wavelength, respectively. **(b)** `Mobile operation': optical images obtained with a lens and imaged on a smartphone, demonstrating the TFMMI's potential for fieldwork.

**CONCLUSIONS**

In summary, we have demonstrated that multimode interference (MMI) in a tapered waveguide enables high-resolution and broadband spectroscopy in a compact, monolithic device. An ultra-narrow spectral resolution down to 40 pm in the visible ($Q = 15,950$) and 10 pm in the IR ($Q = 155,000$) is demonstrated. We experimentally showed that the TFMMI spectrometer operates across an exceptionally large range from 500 nm to 1600 nm ($B = 1.0576$), which is limited only by the transparency of the waveguide material and the sensitivity of the camera. Thus, with suitably chosen materials, this concept can be easily extended into the ultraviolet and mid-infrared spectrum. The technique is also suitable for on-chip tapered multimode waveguides, which could be fabricated in high volume by printing or optical lithography. Such ultra-small and monolithic, high-performance spectrometers are expected to find applications in a wide array of fixed and portable spectroscopy applications in fields ranging from biochemical sensing to the life- and physical sciences.

ACKNOWLEDGEMENTS

F.M. was partly supported by the Chinese Scholarship Council, a research program funded by the Chinese government under award no. 201206470032. D.E. acknowledges support by a Sloan Research Fellowship. R-J.S. was supported in part by the Center for Excitonics, an Energy Frontier Research Center funded by the US Department of Energy, Office of Science, Office of Basic Energy Sciences under award no. DE-SC0001088. E.H.C. was supported by the NASA Office of the Chief Technologist's Space Technology Research Fellowship. T.S. was supported by the Alexander von Humboldt Foundation.

REFERENCES

1. Jackson M.W. *Spectrum of Belief: Joseph von Fraunhofer and the Craft of Precision Optics.* (The MIT Press, Cambridge, MA, 2000).

2. Chan, J. W., Motton D., Rutledge J.C., Keim N.L. & Huser T. Raman spectroscopic analysis of biochemical changes in individual triglyceride-rich Lipoproteins in the pre- and postprandial state. *Anal. Chem.* **77,** 5870-5876 (2005).

3. Fiddler M.N. *et al*. Laser spectroscopy for atmospheric and environmental sensing. *Sensors* **9,** 10447-10512 (2009).

4.  Nitkowski A., Chen L. & Lipson, M. Cavity-enhanced on-chip absorption spectroscopy using microring resonators. *Opt. Express* **16,** 11930-11936 (2008).

5.  Matsumoto, T., Fujita S. & Baba, T, Wavelength demultiplexer consisting of photonic crystal superprism and superlens. *Opt. Express* **13**, 10768-10776 (2005).

6.  Laubscher, M. *et al*. Spectroscopic optical coherence tomography based on wavelength de-multiplexing and smart pixel array detection. *Opt. Commu.* **237,** 275-283 (2004).

7.  Lobiński R., Schaumlöffel D. & Szpunar J. Mass spectrometry in bioinorganic analytical chemistry. *Mass Spectrom. Rev.* **25**, 255-289 (2006).

8.  Apex Technologies Ultrahigh Resolution OSA AP2041B, http://www.apex-t.com/pdf/optical-spectrum-analyzer.pdf

9.  Aragon Photonics BOSA 100 and 200 series, http://aragonphotonics.com/?%20page_id=149

10. Grabarnik S, *et al*. Planar double-grating microspectrometer. *Opt. Express* **15,** 3581- 3588 (2007).

11. Momeni, B., Hosseini, E. S. & Adibi, A. Planar photonic crystal microspectrometers in silicon- nitride for the visible range. *Opt. Express* **17**, 17060-17069 (2009).

12. Pervez N, *et al*. Photonic crystal spectrometer. *Opt. Express* **18,** 8277-8285 (2010).

13. Gan, X., Pervez, N., Kymissis, I., Hatami, F. & Englund, D. A high-resolution spectrometer based on a compact planar two dimensional photonic crystal cavity array. *Appl. Phys. Lett.* **100,** 231104 (2012).

14. Hang Q., Ung B., Syed I., Guo N. & Skorobogatiy M. Photonic bandgap fiber bundle spectrometer. *Appl. Opt.* **49,** 4791-4800 (2010).

15. Redding, B. & Cao, H. Using a multimode fiber as high-resolution, low-loss spectrometer. *Opt. Lett.* **37,** 3384-3386 (2012).

16. Redding, B., Popoff, S. M. & Cao, H. All-fiber spectrometer based on speckle pattern reconstruction. *Opt. Express* **21,** 6584-6600 (2013).

17. Redding, B., Liew, S. F., Sarma, R. & Cao, H. Compact spectrometer based on a disordered photonic chip. Nature Photon. **7,** 746-751 (2013).

18. Velasco A., *et al*. High-resolution Fourier-transform spectrometer chip with microphotonic silicon spiral waveguides. *Opt. Lett.* **38,** 706-708 (2013).

19. Cheben P., *et al*. A high-resolution silicon-on-insulator arrayed waveguide grating microspectrometer with sub-micrometer aperture waveguides. *Opt. Express* **15,** 2299- 2306 (2007).

20. Kyotoku B., Chen L. & Lipson M. Sub-nm resolution cavity enhanced microspectrometer. *Opt. Express* **18**, 102-107 (2010).

21. Schweiger G., Nett R. & Weigel T. Microresonator array for high-resolution spectroscopy. *Opt. Lett.* **32,** 2644-2646 (2007).

22. Xia Z., *et al*. High resolution on-chip spectroscopy based on miniaturized microdonut resonators. *Opt. Express* **19,** 12356-12364 (2011).

23. DeCorby R.G., Ponnampalam N., Epp E., Allen T. & McMullin J.N. Chip-scale spectrometry based on tapered hollow Bragg waveguides. *Opt. Express* **17,** 16632-16645 (2009).

24. Princeton Instruments: Acton Series Monochromators and Spectrographs, http://www.princetoninstruments.com/Uploads/Princeton/Documents/Datasheets/Princeton_Instruments_Acton_Series_N2_4_10_13.pdf

25. Thorlabs AFS105/125Y - Multimode Fiber, 0.22 NA, Low-OH, Ø105 µm Core, Vis-IR

    http://www.thorlabs.com/thorcat/4200/AFS105_125Y-SpecSheet.pdf

26. Brambilla, G., Finazzi V. & Richardson, D. J. Ultra-low-loss optical fiber nanotapers. *Opt. Express* **12,** 2258-2263 (2004).

27. Xu, Z. *et al*. Multimodal multiplex spectroscopy using photonic crystals. *Opt. Express* **11,** 2126–2133 (2003).

28. Birks, T. A. & Li, Y. W. The shape of fiber tapers. *J. Lightwave Technol.* **10,** 432-438 (1992).